\documentclass[usenatbib,referee]{ctextemp_mn2e}

\title[The short-period limit of contact binaries]
{The short-period limit of contact binaries}

\author[Jiang et al.]{Dengkai Jiang$^{1,2}$\thanks{E-mail:
dengkai@ynao.ac.cn}, Zhanwen Han$^{1,2}$, Hongwei Ge$^{1,2}$, Liheng Yang$^{1}$ and Lifang Li$^{1,2}$\\
$^{1}$National Astronomical Observatories, Yunnan Observatory,
Chinese Academy of Sciences, P.O. Box 110,
Kunming,\\
\ \ \ \  \ \ \ \ \ \ \ \ \ \ \ \ \ \ \ \ \ \ \ \ \ \ \ \ \ \ \ \ \
\ \ \ \  Yunnan Province, 650011, P.R. China\\
$^{2}$Key Laboratory for the Structure and Evolution of Celestial
Objects, Chinese Academy of Sciences}
\begin{document}
\input ctextemp_psfig.sty
\date{Accepted .... Received .....; in original form ....}

\pagerange{\pageref{firstpage}--\pageref{lastpage}} \pubyear{2009}

\maketitle

\label{firstpage}

\begin{abstract}
The stability of mass transfer is important in the formation of
contact binaries from detached binaries when the primaries of the
initially detached binaries fill their Roche lobes. Using Eggleton's
stellar evolution code, we investigate the formation and the
short-period limit of contact binaries by considering the effect of
the instability of mass transfer. It is found that with decreasing
initial primary mass from 0.89M$_{\rm \odot}$ to 0.63M$_{\rm
\odot}$, the range of the initial mass ratio decreases for detached
binaries that experience stable mass transfer and evolve into
contact. If the initial primary mass is less than 0.63M$_{\rm
\odot}$, detached binaries would experience dynamically unstable
mass transfer when the primaries of detached binaries fill their
Roche lobes. These systems would evolve into a common envelope
situation and probably then to a complete merger of two components
on a quite short timescale. This results in a low mass limit at
about 0.63M$_{\rm \odot}$ for the primary mass of contact binaries,
which might be a main reason why the period distribution of contact
binaries has a short limit of about 0.22 days. By comparing the
theoretical period distribution of contact binaries with the
observational data, it is found that the observed contact binaries
are above the low mass limit for the primary mass of contact
binaries and no observed contact binaries are below this limit. This
suggests that the short-period limit of contact binaries can be
explained by the instability of the mass transfer that occurs when
the primaries of the initially detached binaries fill their Roche
lobes.

\end{abstract}


\section{Introduction}
Contact binaries are the most common eclipsing binaries in the solar
neighborhood \citep{Shapley 1948}. The period distribution of
contact binaries has a range from 0.22d to more than 100d, but a
strong maximum near 0.37d that is very close to a short-period limit
at about 0.22d \citep{Rucinski 1992, Rucinski 1998, Rucinski 2007,
Paczynski 2006}. The shortest period sky-field contact binary
currently known is GSC 01387-00475 (ASAS 083128+1953.1) which has a
period of 0.2178 d \citep{Rucinski 2007, Rucinski 2008}. The
existence of this system confirms that the short-period limit of
contact binaries at about 0.22d is very sharp and well defined
\citep{Rucinski 2008}, although a shorter contact binary V34 is in
globular cluster 47 Tuc with $P$=0.2155d \citep{Weldrake 2004}. It
is an interesting puzzle to determine why contact binaries have a
very well-defined short-period limit. Moreover, understanding the
reason of this limit helps to improve our understanding of the
theory of stellar and binary evolution with low mass components.

The short-period limit of contact binaries has been investigated by
many authors \citep{Rucinski 1992, Rucinski 2007, Rucinski 2008,
Stepien 2001, Stepien 2006, Paczynski 2006}. \citet{Rucinski 1992}
attempted to explain the short-period limit by the full convection
limit for low-mass stars that the fully convective configuration
leads to a strong limit on the parameters of contact binaries. But
he found that the full convection limit is some distance from the
reddest observed contact systems and he suggested that the full
convection limit is not the main reason of the short-period limit.

A traditional view for the formation of contact binaries is that
they are formed from detached binaries of comparable periods.
Recently, \citet{Stepien 2006} investigated the timescale of
detached binaries for reaching a stage of Roche lobe overflow
(RLOF), and hence for forming contact binaries. He found that for a
detached system with initial mass of the primary of 0.7M$_{\rm
\odot}$ and initial period of 2d, the timescale for reaching RLOF is
greater than the age of the Universe because the angular momentum
loss (AML) timescale increases with decreasing stellar mass.
Therefore, he suggested that the short-period limit of 0.22d of
contact binaries corresponds to a lower limit of around
$1.0-1.2$M$_{\rm \odot}$ for the total mass of the system and is due
to the finite age of the binary population forming contact systems
of several Gyr. However, observations show that there are some
short-period detached (or semidetached) binaries below the total
mass limit, such as OGLE BW3 V38 ($P$=0.198d, Maceroni \&
Montalb\'{a}n 2004), GSC 2314-0530 ($P$=0.192d, Dimitrov 2010), NSVS
01031772 ($P$=0.37d, Lopez-Morales et al. 2006), GU Boo ($P$=0.49d,
Lopez-Morales \& Ribas 2005). The existence of these systems
suggests that the AML rate is higher than that postulated by
\citet{Stepien 2006}, or such low-mass binaries could have very
short periods at birth. Therefore, the short-period limit of contact
binaries could not be explained completely by the finite age of the
binary population forming contact systems of several Gyr suggested
by \citet{Stepien 2006}. It seems that further investigation is
needed in the formation of contact binaries from detached binaries
in order to confirm the reason of the short-period limit.

\citet{Nelson 2001} constructed a large grid of models (0.8M$_{\rm
\odot}$$\leq$ $M_{10}$ $\leq$50M$_{\rm \odot}$) of case A binary
evolution, according to the assumption of conservative evolution.
They showed that in case AR (rapid evolution to contact) and AS
(slow evolution to contact), the secondary expands in response to
the thermal-timescale mass transfer or the nuclear timescale mass
transfer from the primary and fills its own Roche lobe before either
component has left the main sequence (MS). They suggested that this
probably leads to a contact binary. However, in case AD (dynamic
RLOF), the radius of the primary increases faster (or decreases more
slowly) than the Roche lobe when RLOF begins. Mass transfer is
dynamically unstable and quickly accelerates to the dynamic mass
transfer. The secondary can not accrete all the proffered material.
This probably leads to complete engulfment of the secondary,
creating a common envelope binary \citep{Paczynski 1976, Hjellming
1987}. These systems might coalescence on a quite short timescale
\citep{Eggleton 2000, Nelson 2001} and could not form contact
binaries. The instability of mass transfer has been studied in the
past \citep{Hjellming 1987, Ge 2010, Deloye 2010}. \citet{Ge 2010}
show that the instability of mass transfer may occur promptly upon
the primary filling its Roche lobe, if the primary has a surface
convection zone of any significant depth. MS stars with
$M<1.25$M$_{\rm \odot}$ have a convective envelope and the mass of
convective envelope increases with decreasing mass of stars
\citep{Hurley 2000}. This means that the instability of mass
transfer needs to be taken into account in investigating the
formation of contact binaries from detached binaries with less
massive components.

The purpose of this study is to investigate the period limit of
contact binaries. Employing the Eggleton stellar evolution code, we
construct a grid of binary models for a Population I metallicity
$Z$= 0.02. We compare the observational data with the theoretical
period distribution of contact binaries, and find that the
short-period limit could be explained by the instability of mass
transfer that occurs when the primaries of the initially detached
binaries fill their Roche lobes.

\section{The short-period limit of contact binaries}

In the formation of contact binary from detached binary, the primary
of detached binary fills its Roche lobe and transfers some of its
mass to the secondary. The secondary would expand to fill its Roche
lobe and the system come into contact. To determine whether the mass
transfer rate reaches the dynamic timescale before the secondary
fills its Roche lobe, it is necessary to perform detailed binary
evolution calculations. Here we use Eggleton's stellar evolution
code to study the instability of mass transfer in the formation of
contact binaries from detached binaries. This code was originally
developed by \citet{Eggleton 1971, Eggleton 1972} and
\citet{Eggleton 1973} and has been updated with the latest input
physics during the last four decades (e.g. Han, Podsiadlowski \&
Eggleton 1994; Pols et al. 1995, 1998; Nelson \& Eggleton 2001;
Eggleton \& Kiseleva-Eggleton 2002). The current code includes a
model of dynamo-driven mass loss, magnetic braking, and tidal
friction to the evolution of stars with cool convective envelopes,
and the simplification is considered that only the primary is
subject to these nonconservative effects \citep{Eggleton 2002}.
Considering these nonconservative effects, we construct a grid of
stellar evolutionary models that covers the following ranges of
initial primary mass $M_{10}$ and initial mass ratio
($q_0=M_{20}/M_{10}$):
\begin{equation}
{\rm log} M_{10} = -0.25, -0.2,-0.19, -0.15, -0.1, -0.05,
\end{equation}
\begin{equation}
{\rm log} (1/q_{0}) = 0.25, 0.2, 0.15, 0.1, 0.05, 0.02.
\end{equation}
We present one initial period, log ($P_0$/$P_{\rm ZAMS}$)=0.5, where
$P_{\rm ZAMS}$ is the period at which the initially more massive
component would just fill its Roche lobe on the zero-age main
sequence \citep{Nelson 2001}.

\begin{figure*}
\centerline{\psfig{figure=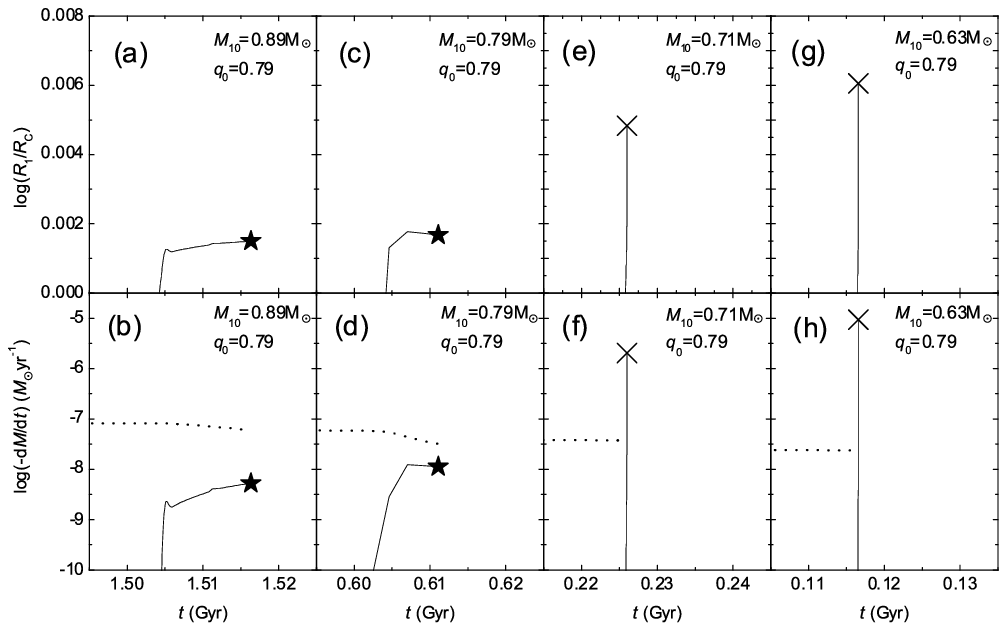,width=15.0cm}} \caption{Four
examples of binary evolution calculations. The solid lines show the
evolutionary tracks of the primaries after the primaries fill their
Roche lobes. The radius ratio of the primary ($R_1/R_c$) is shown in
panels (a), (c), (e), (g) and the mass transfer rate (${\rm d}M/{\rm
d}t$) is shown in panels (b), (d), (f), (h). The dotted lines
represent 10$M/t_{\rm KH}$ that is used to determined whether the
mass transfer is dynamic. Filled Stars indicate the position where
the secondary fills its Roche lobe and the system evolves into
contact. Crosses represent that mass transfer is dynamically
unstable. The initial binary parameters are given in each panel.}
\label{fig2}
\end{figure*}

In Fig.1, we present four examples (solid lines) of our binary
evolution calculations with log(1/$q_{0}$)=0.1 ($q_0$=0.79) after
the primaries fill their Roche lobes. It shows the radius ratio of
the primary ($R_1/R_{\rm c}$, where $R_1$ and $R_{\rm c}$ are the
radius of the primary and the Roche lobe radius of the primary) and
the mass transfer rate (d$M$/d$t$). Figs 1(a) and (b) represent the
evolution of a binary system with an initial mass of the primary of
log$M_{10}= -0.05$ ($M_{10}$=0.89M$_{\rm \odot}$). The primary fills
its Roche lobe on the MS which results in case A RLOF. The ratio of
$R_1/R_{\rm c}$ increases slower after this value is greater than
0.0012 as shown in Fig 1(a). Then, the mass transfer is stable and
the mass-transfer rate does not increase quickly as shown in Fig
1(b). \citet{Nelson 2001} suggested that the mass transfer is
determined to be dynamic when the mass-transfer is greater than
$10M/t_{\rm KH}$ ($t_{\rm KH}$ is the thermal or Kelvin-Helmholtz
timescale), which is shown as dotted line in Fig 1. The
mass-transfer rate of this system is not beyond the dotted line
before the secondary fills its Roche lobe and this system evolves
into contact shown as filled star. Figs 1(c) and (d) show another
example for an initial system with log$M_{10}= -0.1$
($M_{10}$=0.79M$_{\rm \odot}$). The binary evolves in a similar way
as in the previous example and the main difference between this
example and the previous one is the ratio of $R_1/R_{\rm c}$ roughly
constant after the value is greater than 0.0017.

Figs 1(e) and (f) represent the third example for initial system
with log$M_{10}= -0.15$ ($M_{10}$=0.71M$_{\rm \odot}$). Fig 1(g) and
(h) represent the last example for initial system with log$M_{10}=
-0.2$ ($M_{10}$=0.63M$_{\rm \odot}$). In these two systems, the
radius ratio of the primary ($R_1/R_{\rm c}$) increases quickly
after the onset of RLOF. This leads to the mass-transfer rate that
increases sharply, and is beyond the dotted line. Hence the mass
transfer becomes dynamically unstable. Such systems might evolve
into a common envelope situation and probably coalescence on a quite
short timescale \citep{Eggleton 2000, Nelson 2001}.

\begin{figure}
\centerline{\psfig{figure=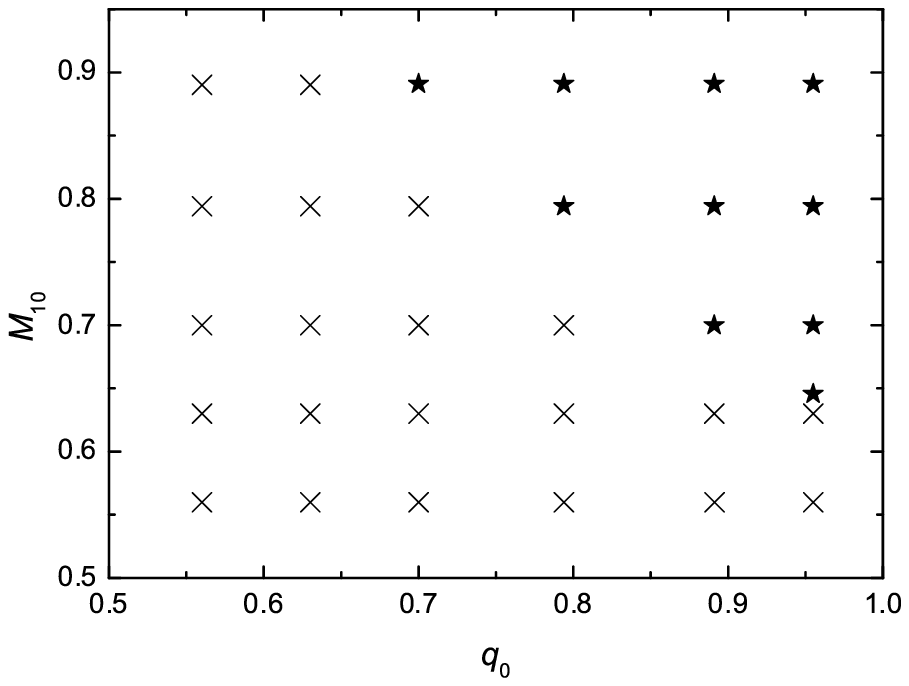,width=7.5cm}} \caption{Final
outcomes of the binary evolution calculations in the initial mass
ratio-primary mass ($q_0, M_{10}$) plane. Crosses denote the systems
that experience the dynamical instability and filled stars indicate
contact binaries are formed without experiencing the dynamical
instability as shown in Fig 1. } \label{fig1}
\end{figure}

Fig. 2 summarizes the final outcome of binary evolution calculations
in the initial mass ratio-primary mass ($q_0, M_{10}$) plane.
Crosses show systems that are unstable to dynamical mass transfer
and filled stars indicate systems that do not experience the
dynamical instability and evolve into contact. It is seen from Fig.
2 that with decreasing initial mass of the primaries from
0.89M$_{\rm \odot}$ to 0.63M$_{\rm \odot}$, the range of the initial
mass ratio decreases for detached binaries that experience stable
mass transfer and evolve into contact, and the range of initial mass
ratio increases for detached binaries such that the mass transfer is
on the dynamical timescale. If the initial primary mass is less than
0.63M$_{\rm \odot}$,  the initially detached binaries would suffer
dynamically unstable mass transfer.

\begin{table*}
\begin{footnotesize}
Table~1.\hspace{4pt} The physical parameters of some binaries with short period.\\
\begin{minipage}{12cm}
\begin{tabular}{l|ccccccc}
\hline\hline\ {Stars}&{Type}&{$P$}&{$M_{1}$}&{$M_{2}$}
&{$R_{1}$}&{$R_{2}$}&{References}\\
&&{(d)}&{(M$_{\rm \odot}$)}&{(M$_{\rm \odot}$)}
&{(R$_{\rm \odot}$)}&{(R$_{\rm \odot}$)}\\
\hline

GSC 01387-00475& C &0.2178  &0.638 &0.302   &-        &-       &(1)\\
CC Com         & C &0.221   &0.79  &0.41    &0.73     &0.54    &(2)\\
V523 Cas       & C &0.2337  &0.75  &0.38    &0.74     &0.55    &(3)\\
BI Vul         & C &0.2518  &0.86  &0.59    &0.82     &0.7     &(2)\\
VZ Psc         & C &0.2612  &0.79  &0.72    &0.77     &0.74    &(2)\\
FS Cra         & C &0.2636  &0.86  &0.65    &0.82     &0.73    &(2)\\
FG Sct         & C &0.2706  &0.87  &0.68    &0.83     &0.73    &(2)\\
XY Leo         & C &0.2841  &0.82  &0.50    &0.86     &0.69    &(4)\\
TZ Boo         & C &0.2976  &0.72  &0.11    &0.97     &0.43    &(4)\\
V829 Her       & C &0.3582  &0.856 &0.372   &1.058    &0.711   &(5)\\
FI Boo         & C &0.39    &0.82  &0.31    &1.1      &0.71    &(6)\\
BH Cas         & C &0.4059  &0.73  &0.35    &1.09     &0.78    &(7)\\

GSC 2314-0530  & SD &0.192  &0.51  &0.26    &0.55     &0.29    &(8)\\
OGLE BW3 V38   & D &0.198  &0.44  &0.41    &0.51     &0.44    &(9)\\

NSVS 01031772  & D &0.37  &0.54  &0.50    &0.53     &0.51    &(10)\\
NSVS 07453183  & D &0.37  &0.73  &0.68    &0.79     &0.72    &(11)\\
V405 And       & D &0.465 &0.49  &0.21    &0.78     &0.23    &(12)\\
GU Boo         & D &0.49  &0.61  &0.60    &0.62     &0.62    &(13)\\
NSVS 06507557  & D &0.51  &0.65  &0.28    &0.60     &0.44    &(14)\\
2MASS 04463285+1901432   & D &0.62  &0.47  &0.19    &0.56     &0.21    &(15)\\
YY Gem         & D &0.81  &0.60  &0.60    &0.62     &0.62    &(16)\\

\hline
\end{tabular}
\end{minipage}
\end{footnotesize}\\
{Columns: Stars-name of star; Type-type of binary configuration(C =
contact, SD = semidetached, D = detached); $P$-orbital period;
$M_1$-mass of the primary; $M_2$-mass of the secondary; $R_1$-radius
of the primary;$R_2$-radius of the secondary.\\
References in Table 1: (1)Rucinski (2008); (2)Maceroni \& van¡¯t
Veer (1996); (3)Zhang \& Zhang (2004); (4)Yakut \& Eggleton (2005);
(5)Gazeas et al. (2005); (6)Terrell et al. (2006); (7)Zo{\l}a,
Niarchos \& Dapergolas (2001); (8) Dimitrov \& Kjurkchieva (2010);
(9) Maceroni \& Montalb\'{a}n (2004); (10) Lopez-Morales et al.
(2006); (11) Coughlin \& Shaw (2007); (12) Vida et al. (2009); (13)
Lopez-Morales \& Ribas (2005); (14) Cakirly \& Ibanoglu (2010); (15)
Hebb et al.(2006); (16) Bopp (1974), Torres \& Ribas (2002).}
\end{table*}

\begin{figure}
\centerline{\psfig{figure=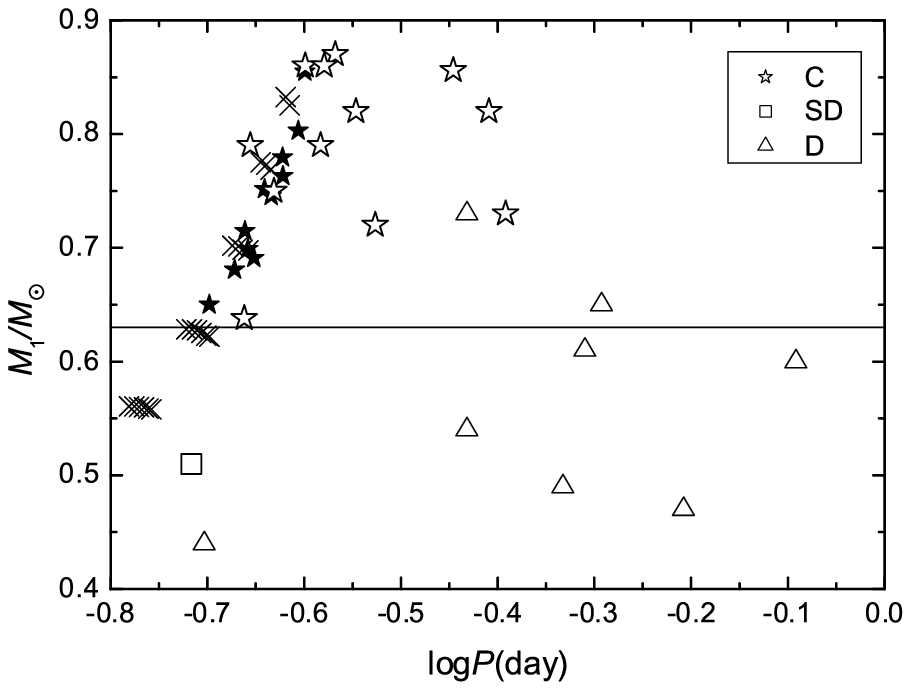,width=7.5cm}} \caption{The
theoretical distribution of the period and the primary mass of
binaries when they evolve into contact (filled stars) and that of
binaries that experience the dynamical instability (crosses). Open
stars show the position of observed contact binaries, open square
and open triangles represent observed semidetached binary and
observed detached binaries, respectively.} \label{fig2}
\end{figure}

Fig. 3 shows the theoretical distribution of the period and the
primary mass of the binaries when they evolve into contact (filled
stars) and that of binaries that experience the dynamical
instability (crosses). It is seen in Fig. 3 that the theoretical
period of binaries that experience the dynamical instability could
be as short as 0.165d (log$P=-0.781$). The theoretical period of
contact binaries has a low limit at about 0.20d (log$P=-0.698$),
which results from the low mass limit at about 0.63M$_{\rm \odot}$
for the primary mass of contact binaries (solid line). This suggests
that the distribution of the period of contact binaries depends on
the formation of contact binaries, and then on the stability of mass
transfer when the primaries of detached binaries fill their Roche
lobes.

In addition, we collected the physical parameters of some binaries
with short period where both components are MS stars from literature
(listed in Table 1). These observed systems are also plotted in Fig.
3 with open symbols (open stars = contact binaries, open square =
semidetached binary, open triangles = detached binaries). As seen
from Fig. 3, the observed contact binaries are above the low mass
limit for the primary mass of contact binaries. There are no
observed contact binaries below this limit. We also note that one
semidetached binary and some detached binaries are below this limit.

\begin{figure}
\centerline{\psfig{figure=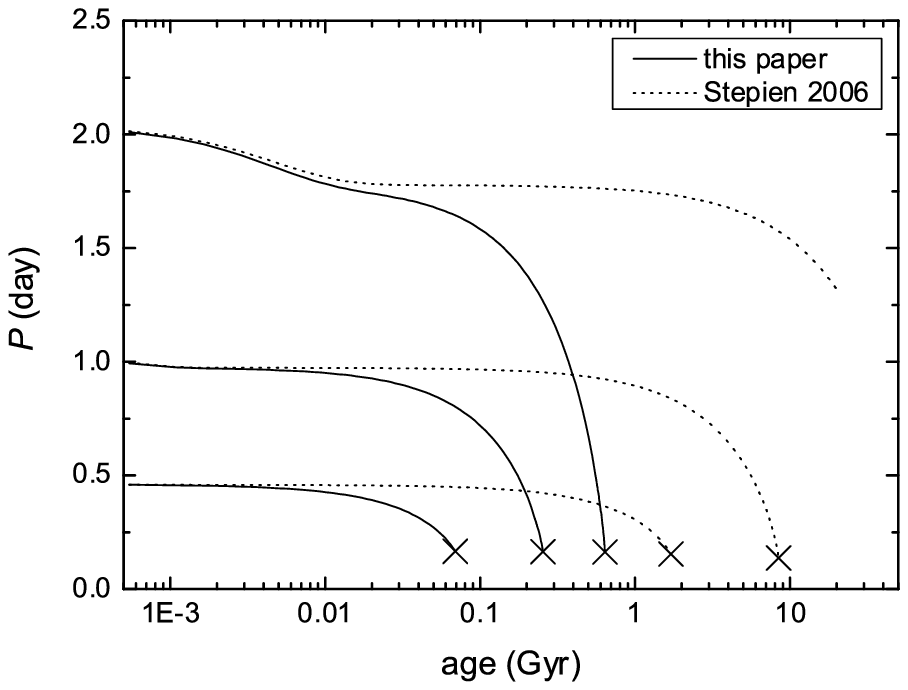,width=7.5cm}} \caption{A
comparison of the orbital evolution of initially detached binaries
in our model (solid lines) with that in the model based on the AML
rate given by \citet{Stepien 2006} (dotted lines) for systems with
log $M_{10} =-0.25$, log $1/q_{0}$ = 0.25, and $P_0 = 0.46, 1.0,
2.0$d. Crosses denote the systems that experience the dynamical
instability.} \label{fig2}
\end{figure}

We compare the orbital evolution of the initially detached binaries
in our model with that in the model based on the AML rate given by
\citet{Stepien 2006}. They are shown in Fig. 4 for the systems with
log $M_{10} =-0.25$, log $1/q_{0}$ = 0.25, and $P_0 = 0.46, 1.0,
2.0$d. It is seen from Fig. 4 that the initially detached binaries
in our model (solid lines) reach RLOF more rapidly than those in
model based on St\c epie\'n's AML rate (dotted lines). For example,
a system with an initial period $P_0 = 1.0$d in the model adopting
St\c epie\'n's AML rate spends a much longer time ($\sim8$Gyr) to
reach RLOF than that in our model ($\sim0.25$Gyr). This suggests
that the AML rate assumed by \citet{Stepien 2006} is much lower than
the rate used in this study in \citet{Eggleton 2002}.

\section{Discussion and conclusions}

In this paper, we investigated the short-period limit of contact
binaries by using the Eggleton stellar code with considering the
effect of the instability of mass transfer that occurs when the
primaries of detached binaries fill their Roche lobes.

\citet{Stepien 2006} suggested that the short-period limit might be
caused by the fact that the initially detached binaries with
low-mass components do not have time to reach RLOF even within the
age of the Universe since the AML timescale increases with
decreasing stellar mass. However, this could not explain the
existence of the short-period low-mass binary systems, such as V405
And. This suggests that the AML rate is underestimated by
\citet{Stepien 2006}, or some low-mass binaries could have a very
short orbital period at their birth. We found that contact binaries
have a low mass limit at about 0.63M$_{\rm \odot}$ for the primary
mass due to the instability of the mass transfer when the primaries
of the initially detached binaries fill their Roche lobes. This
suggests that the formation of contact binaries depends on the
stability of mass transfer when the primaries of detached binaries
fill their Roche lobes.

The distribution of the period of contact binaries depends on the
formation of contact binaries, and then on the stability of mass
transfer when the primaries of detached binaries fill their Roche
lobes. By comparing the theoretical period distribution of contact
binaries with the observational data as shown in Fig 3, it is found
that the observed contact binaries are above the low mass limit for
the primary mass of contact binaries and no observed contact
binaries are below this limit. This means that the observed contact
binaries might be formed from detached binaries that experience
stable mass transfer and the short-period limit of contact binaries
can be explained by the instability of the mass transfer that occurs
when the primaries of detached binaries fill their Roche lobes. We
also note that one semidetached binary (GSC 2314-0530) and some
detached binaries are below this limit. These systems would
experience unstable mass transfer and coalescence on a quite short
timescale. This suggests that GSC 2314-0530 might be still a
detached binary as suggested by \citet{Norton 2011}, but the primary
might be very close to its Roche lobe.

Our study showed that contact binaries have a low mass limit at
about 0.63M$_{\rm \odot}$ for the primary mass due to the
instability of the mass transfer. It should be noted that the range
of the initial mass ratio is very narrow for systems ($M_{10} \leq
0.7$M$_{\rm \odot}$) that do not experience dynamically unstable
mass transfer. This might be the reason why there is only one
contact binary GSC 01387-004750 \citep{Rucinski 2008} where the
primary mass is smaller than 0.7M$_{\rm \odot}$. \citet{Tout 1997}
and \citet{Hurley 2002} suggested that mass transfer to a component
proceeds dynamically if the primaries are low-mass MS stars ($M \leq
0.7$M$_{\rm \odot}$) that are deeply convective. Our result is not
much different from the suggestion given by \citet{Tout 1997} and
\citet{Hurley 2002}.

In this paper we only presented a grid of stellar evolutionary
models with one initial period to study the stability of mass
transfer in the formation of contact binaries. But many parameters,
such as the initial period, the metallicity, the mass loss, etc.,
are related to the stability of mass transfer, which is also
affected by the detailed process of mass transfer. The effects of
these parameters need to be taken into account in the future study
of the stability of mass transfer and the formation of contact
binaries.

\section*{ACKNOWLEDGEMENTS}
It is a pleasure to thank the referee prof. P.P. Eggleton for his
valuable suggestions and comments, which improve the paper greatly.
This work was supported by the Chinese Natural Science Foundation
(10773026, 10821061, 11073049, 2007CB815406, 11033008 and 11103073),
the Foundation of Chinese Academy of Sciences (KJCX2-YW-T24).

\bsp

\label{lastpage}

\end{document}